\documentclass[reprint, superscriptaddress, preprintnumbers, amsmath, amssymb, aps, showpacs, showkeys, prl]{revtex4-1}

\usepackage{graphicx}
\usepackage{dcolumn}
\usepackage{bm}
\usepackage{color}

\begin{document}

\title{Nonlinearity in the dark: Broadband terahertz generation with extremely high efficiency}

\author{Ming Fang}
\affiliation{Ames Laboratory---U.S.~DOE and Department of Physics and Astronomy, Iowa State University, Ames, Iowa 50011, USA}
\affiliation{Key Laboratory of Intelligent Computing and Signal Processing, Ministry of Education, Anhui University, Hefei 230039, China}

\author{Nian-Hai Shen}
\email[]{nhshen@ameslab.gov}
\affiliation{Ames Laboratory---U.S.~DOE and Department of Physics and Astronomy, Iowa State University, Ames, Iowa 50011, USA}

\author{Wei E.I. Sha}
\affiliation{Key Laboratory of Micro-nano Electronic Devices and Smart Systems of Zhejiang Province, College of Information Science and Electronic Engineering, Zhejiang University, Hangzhou 310027, China}

\author{Zhixiang Huang}
\email[]{zxhuang@ahu.edu.cn}
\affiliation{Key Laboratory of Intelligent Computing and Signal Processing, Ministry of Education, Anhui University, Hefei 230039, China}

\author{Thomas Koschny}
\affiliation{Ames Laboratory---U.S.~DOE and Department of Physics and Astronomy, Iowa State University, Ames, Iowa 50011, USA}

\author{Costas M. Soukoulis}
\affiliation{Ames Laboratory---U.S.~DOE and Department of Physics and Astronomy, Iowa State University, Ames, Iowa 50011, USA}
\affiliation{Institute of Electronic Structure and Lasers (IESL), FORTH, 71110 Heraklion, Crete, Greece}

\date{\today}

\begin{abstract}
Plasmonic metamaterials and metasurfaces offer new opportunities in developing high performance terahertz emitters/detectors beyond the limitations of conventional nonlinear materials. However, simple meta-atoms for second-order nonlinear applications encounter fundamental trade-offs in the necessary symmetry-breaking and local-field enhancement due to radiation damping that is inherent to the operating resonant mode and cannot be controlled separately. Here we present a novel concept that eliminates this restriction obstructing the improvement of terahertz generation efficiency in nonlinear metasurfaces based on metallic nano-resonators. This is achieved by combing a resonant dark-state metasurface, which locally drives nonlinear SRRs in the near field, with a specific spatial symmetry that enables destructive interference of the radiating linear moments of the SRRs, and perfect absorption via simultaneous electric and magnetic critical coupling of the pump radiation to the dark mode. Our proposal allows eliminating linear radiation damping, while maintaining constructive interference and effective radiation of the nonlinear components. We numerically demonstrate a giant second-order nonlinear susceptibility $\sim10^{-11}$ m/V, one order improvement compared to previously reported split-ring-resonator metasurface, and correspondingly, two orders of magnitude enhanced terahertz energy extraction should be expected with our configuration under the same conditions. Our study offers a paradigm of high efficiency tunable nonlinear meta-devices and paves the way to revolutionary terahertz technologies and optoelectronic nano-circuitry.
\end{abstract}

\maketitle


Following the emergence of quantum-cascade lasers \cite{Kohler2002, Williams2007, Scalari2009} and ultrafast photoconductive switches \cite{Tonouchi2007}, terahertz (THz), a historically mysterious electromagnetic spectrum bridging microwaves and optics, in the past two decades has gained significant progresses in technologies, leading to tremendous transformative applications in sensing, imaging and communication etc \cite{Tonouchi2007, Mittleman2003, Song2011, Siegel2002, Jansen2010, Federici2010}. The blooming of THz technologies and related applications at current stage demand urgently efficient and compact THz emitters/detectors, yet the development of which remains challenging. Many state-of-the-art THz sources are based on nonlinear optical rectification or difference-frequency generation (DFG) in inorganic crystals, which suffer from the drawbacks of either narrow bandwidth, subtle phase-matching limitations, inevitable spectrum gaps or low emission intensity \cite{Kohler2002, Williams2007, Scalari2009, Lewis2014, Ferguson2002, Belkin2007, Yang2007}. Surpassing natural materials, the freedom of designing our own subwavelength-scale “atoms” as building blocks of metamaterials or metasurfaces has shown great flexibilities in achieving versatile functions in different disciplines \cite{Engheta2007, Soukoulis2011, Koenderink2015, Chen2016, Grady2013, Kravets2013, Meinzer2014, Kildishev2013, Yu2011, Minovich2015, Zheludev2012, Baev2015, Yu2014}. The coexistence of resonant nonlinearity and local field enhancement in properly designed meta-atoms \cite{Minovich2015, Linden2012, Ginzburg2012, Lapine2014, Husu2012, Wolf2015, Kauranen2012} open up an alternative route for high-efficiency nonlinear devices that eliminates those common restrictions. Screening effects of the free-electron gas at the surface of metallic nanoparticles lead to strong second-order nonlinearity in their optical electric susceptibility, which contribute constructively for the area of the metasurface with the specifically designed non-symmetric shape of meta-atoms \cite{Ciraci2012, Zeng2009}. Very recently, a metasurface of 40 nm-thick split-ring resonators (SRRs) was demonstrated to generate broadband gapless THz emission (up to 4 THz) with the conversion efficiency close to 0.2 mm-thick optimal ZnTe crystals \cite{Luo2014}. Despite of being a seminal progress toward compact tailorable THz sources even subwavelength all-optical active devices benefiting next generation computations and communications, the nonlinear optical response reported in this work is still fairly weak. Therefore, a fundamental improvement in the nonlinear efficiency with metasurface-based devices is critical.

In this Letter, we reveal a fundamental limit of conversion efficiency in nonlinear metasurfaces, and accordingly, we propose a dark-mode-assisted perfect absorbing metasurface for THz generation that allows to circumvent it. We demonstrate broadband THz emissions of the metasurface with giant efficiency, about two orders stronger compared to that of conventional SRR metasurface \cite{Luo2014}. This opens up the opportunities for not only revolutionary THz technologies but also high efficiency on-chip nonlinear nano-optical devices.

Analyzing the performance of the SRR metasurface and we find that the major obstacles limiting the nonlinear conversion efficiency are dissipative and radiative damping due to linear scattering. These two loss channels essentially govern the amplitude of the local near field, which in turn reduces the possible THz emission intensity. For the second-order nonlinearity, the ratio of generated THz signal to scattered pump radiation is directly proportional to the local near-field amplitude, thus the linear radiation damping reduces the THz output. While the dissipative loss is hard to minimize and essentially given by the available constituent materials, we, therefore, target the radiative loss channel. Theoretically, we should be able to tackle these restrictions and approach the critical limit in nonlinear efficiency of a metasurface: a properly designed \textit{perfect absorbing} metasurface is applied to eliminate the linear radiative damping and a low-loss dielectric dark element can be introduced for high quality-factor strong local field enhancement to drive the plasmonic nanoparticles, which are generating the nonlinear signals.

\begin{figure}[htb]
\includegraphics[width= 8 cm]{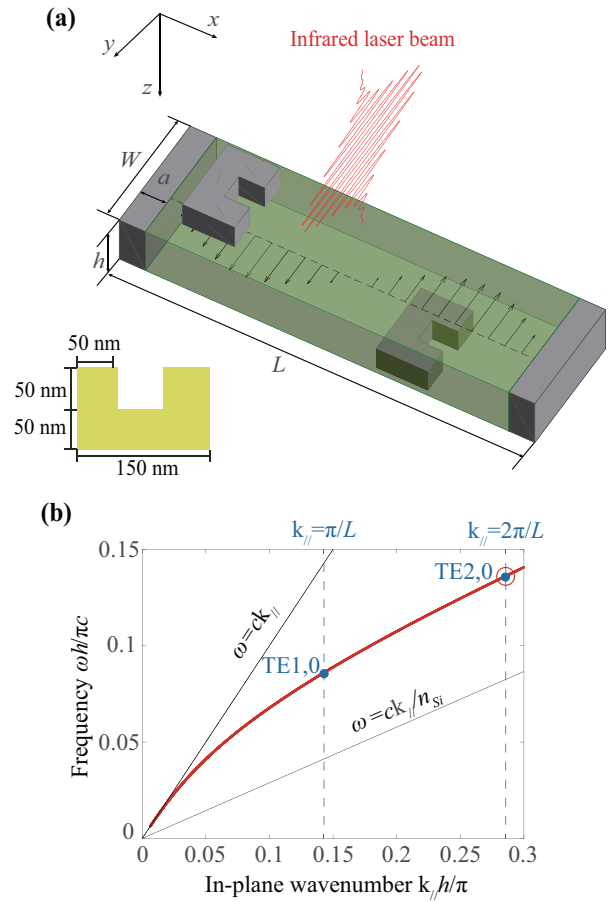}
\caption{\label{fig1} Designed dark-mode THz emitter and principle of operation. (a) Schematics of the meta-atom of designed metasurface THz emitter. (b) Dispersion diagram of the system. The solid (red) line shows the lowest TE-branch and vertical dashed (blue) lines indicate the quantization of eigenmodes. The blue dots are the quantized eigenstates of lowest TE branch, and the red circle indicates the TE$_{2,0}$ mode adopted in this work.}
\end{figure}

In order to implement the concept, we tried to separate the resonant local field enhancement from the current topology necessary for non-vanishing second-order nonlinearity. Instead of driving the SRR directly, we used a resonant dark bound state in a silicon film with a subwavelength thickness to create local field enhancement and the strong evanescent surface fields of this quasi-surface mode then drive the SRRs. The low-loss dielectric dark state allows us to concentrate significant amounts of electromagnetic field energy in the metasurface and to create a high \textit{Q}-factor with strong local field enhancement. Figure~\ref{fig1}(a) presents the schematic unit of our designed metasurface THz emitter. Two gold SRRs (thickness {\color{red}{$t_{\rm{m}}$}}), oriented in the same direction, asymmetrically sit on the two sides of a thin Si slab (thickness \textit{h}), which is periodically punctuated by metal walls (gold, width 2$a$). The two SRRs are not only the key element for THz emission via DFG, but also serve as a bridge coupling the incident wave into the dark element (Si slab). The introduction of metal walls is to quantize the modes of the dielectric slab achieving a set of discrete resonant dark states (see Fig.~\ref{fig1}(b)). Depending on the branch of the dispersion, we may choose one desired mode frequency with the specific corresponding spatial distribution of the composite system for operation. Here, for simplicity, we select the second quantized mode on the lowest branch of dispersion for the dielectric slab, that is, TE$_{2,0}$ mode (marked by red circle in Fig.~\ref{fig1}(b)), which gives an anti-symmetric profile of the electric field as shown in Fig.~\ref{fig1}(a), a dark mode not being able to directly couple to the incident plane wave. The geometric parameters of the unit are designed to have the length $L = 800$ nm, width $W = 240$ nm and thickness $h = 100$ nm, so the operation wavelength is tuned within the typical high-speed telecommunications range at around $1.5~\mu$m. The thickness of SRRs is set as {\color{red}$t_{\rm{m}} = 50$} nm.

\begin{figure*}[htb]
\includegraphics[width= 14 cm]{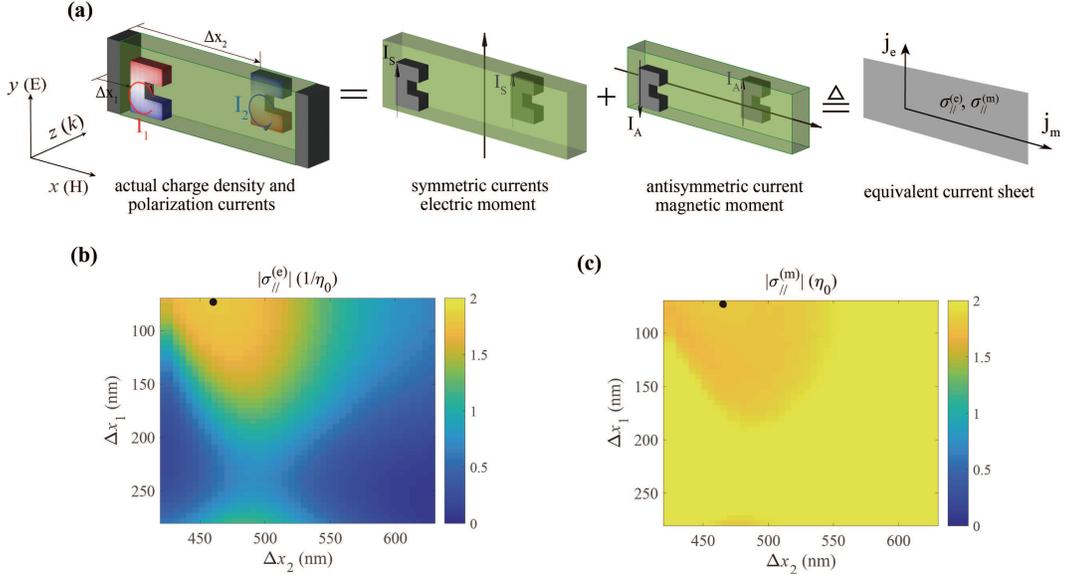}
\caption{\label{fig2} Equivalent current sheet. (a) Polarization currents on two SRRs ($\textbf{I}_1$ and $\textbf{I}_2$) can be decomposed to the set of symmetric ($\textbf{I}_{\rm{S}}$) and antisymmetric ($\textbf{I}_{\rm{A}}$) currents, equivalently described by electric ($\sigma_{\rm{s}}^{\rm{(e)}}$) and magnetic ($\sigma_{\rm{s}}^{\rm{(m)}}$) sheet conductivities. (b) and (c) Maps of $\sigma_{\rm{s}}^{\rm{(e)}}$ and $\sigma_{\rm{s}}^{\rm{(m)}}$ in dependence of positions of the two SRRs. Black dots indicate the optimized case of the system showing the highest absorption.}
\end{figure*}

Coupling of the pump radiation to the resonant dark state can be achieved via the non-resonant weak SRR scatterers, which is designed to simultaneously couple both the electric and magnetic component of the pump to the resonant dark state by creating equivalent electric and magnetic effective surface susceptibilities in the metasurface. In-phase linear currents in the two SRRs created electric moments and anti-phase linear currents in SRRs on opposing sides of the film created magnetic moments and absolute and relative strength of both can be controlled by the symmetry and lateral positioning of the SRRs. By shifting the positions of the two SRRs, i.e., $\Delta x_1$ and $\Delta x_2$, the energy of excitation exchanged to the dark state indeed can be controlled purposely. Upon the excitation of the TE$_{2,0}$ mode in the Si slab, out-of-phase polarization currents $\textbf{I}_1$ and $\textbf{I}_2$ are induced in the two SRRs, which can be decomposed to symmetric (corresponding to an electric moment along \textit{y}-axis) and anti-symmetric (corresponding to a magnetic moment along \textit{x}-axis) polarization sets, respectively, as shown in Fig.~\ref{fig2}(a). Balancing the electric and magnetic coupling allowed for impedance match, hence, eliminating the “reflection” of the pump energy. Tuning the coupling strength to a critical case also allowed us to further eliminate the linear “transmitted” field, thus leading to a complete channeling of the incident pump energy into the dark resonant state and dissipative loss in the metal structures. Consequently, the proposed metasurface can be characterized by the set of electric and magnetic sheet conductivities, i.e., $\sigma_{\rm{s}}^{\rm{(e)}}$ and $\sigma_{\rm{s}}^{\rm{(m)}}$, which are directly connected with the scattering properties, i.e., reflection, \textit{R} and transmission, \textit{T}, of the metasurface in expressions of: $R=2(\zeta\sigma_{\rm{s}}^{\rm{(e)}}-\zeta^{-1}\sigma_{\rm{s}}^{\rm{(m)}})/(4+2\zeta\sigma_{\rm{s}}^{\rm{(e)}}+2\zeta^{-1}\sigma_{\rm{s}}^{\rm{(m)}}+\sigma_{\rm{s}}^{\rm{(e)}}\sigma_{\rm{s}}^{\rm{(m)}})$ and $T=(4-\sigma_{\rm{s}}^{\rm{(e)}}\sigma_{\rm{s}}^{\rm{(m)}})/(4+2\zeta\sigma_{\rm{s}}^{\rm{(e)}}+2\zeta^{-1}\sigma_{\rm{s}}^{\rm{(m)}}+\sigma_{\rm{s}}^{\rm{(e)}}\sigma_{\rm{s}}^{\rm{(m)}})$, where $\zeta$ is the wave impedance \cite{Tassin2012}. An ideal perfect absorbing sheet should fulfill the condition $\zeta\sigma_{\rm{s}}^{\rm{(e)}}=\zeta^{-1}\sigma_{\rm{s}}^{\rm{(m)}}=2$. For our design, we calculated the sheet conductivities in dependence of ($\Delta x_1$,$\Delta x_2$) with a scanning step size of 5 nm, balancing the simulation resolution and cost. The results are shown in Figs.~\ref{fig2}(b) and ~\ref{fig2}(c), from which, the optimal case may be picked at $\Delta x_1\cong70$ nm and $\Delta x_2\cong470$ nm with $\zeta|\sigma_{\rm{s}}^{\rm{(e)}}|=\zeta^{-1}|\sigma_{\rm{s}}^{\rm{(m)}}|=1.8$ and the corresponding absorption reaches as high as 94\% (see below in Fig.~\ref{fig3}).

Depending on the relative positions of the two SRRs, tunable linear radiation damping can be achieved as shown in Fig.~\ref{fig3}(a). For the optimized configuration with \textit{nearly perfect absorption}, Fig.~\ref{fig3}(b) presents the calculated transmission (\textit{T}), reflection (\textit{R}) and absorption (\textit{A}) spectra for the optimized configuration, with the shaded area indicating the dissipation in the two SRRs. In Fig.~\ref{fig3}(c), we show the field distributions, i.e., \textit{y}-component of the electric field, $E_y$, indicated in color and the magnetic field, $\bf{H}$, marked with black arrows, in \textit{x-z} plane perpendicularly across the center of Si slab at $y = 0$. From Fig.~\ref{fig3}(c) we see that the high quality-factor TE$_{2,0}$ mode of the Si slab is strongly excited, featuring an anti-symmetric electric field profile, leading to a dramatic field enhancement around SRRs. Considering the metal walls always coincide with a field minimum of the dark mode, most of the incident energy will be dissipated in SRRs and converted to nonlinear signals. We need to point out that the dimensions of the SRRs are not critical, provided that the requirements of proper dark-mode frequency and perfect absorption for eliminating the linear radiation are fulfilled.

\begin{figure}[htb]
\includegraphics[width = 8.5 cm]{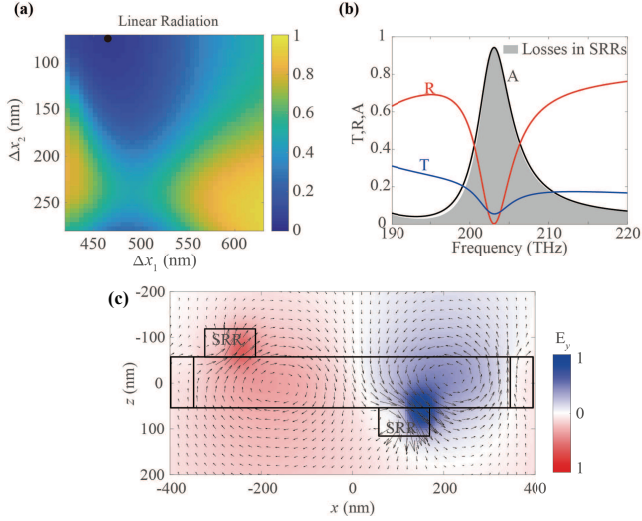}
\caption{\label{fig3} Scattering properties of the metasurface. (a) Normalized linear radiation damping as each scatterer is shifted along the unit cell, black dots indicate the optimized case. (b) Transmission, reflection and absorption spectra for the optimized metasurface. The gray-shaded region shows the dissipation in two SRRs. (c) Color plot shows the $y$-component of the electric field, $E_y$, and black arrows indicate the magnetic field, $\bf{H}$, in $x$-$z$ plane across the center of the meta-atom.}
\end{figure}

For the nonlinear process, THz radiation, we can choose to re-use the two SRRs, which are anyway necessary for providing nonlinearity. Since the desired DFG (or also second-harmonic generation) are second order in the electric filed $\bf{E}$, the sign of the currents does not matter and the nonlinear radiation of both SRRs adds constructively in the far field. The linear radiation, however, is opposite in phase and cancels in the far field, thus completely eliminating linear radiation damping. Since radiation damping of the scattered linear fields, after local dissipation, is the second biggest loss channel in the conversion of pump energy to THz radiation via DFG, this cancellation can significantly increase the efficiency. Consequently, the two SRRs, driven by the TE$_{2,0}$ mode of the slab, show same second-order nonlinear dynamic but opposite linear electric and magnetic dipole moments, resulting in linear radiation cancellation and significant enhancement of THz radiation. By illuminating the designed metasurface with a laser pulse at the resonant frequency of the dark mode, strong THz emission from the SRRs will be observed due to DFG arising from the electron gas in the metal of the SRRs, which can be described by the hydrodynamic model \cite{Ciraci2012, Zeng2009} (see Supplemental Material for details). In the model, the convection term plays the crucial role for the second-order nonlinearities, qualitatively contributing in the format as the product of current and accumulated charge density on the metal surface.

\begin{figure}[htb]
\includegraphics[width= 8.5 cm]{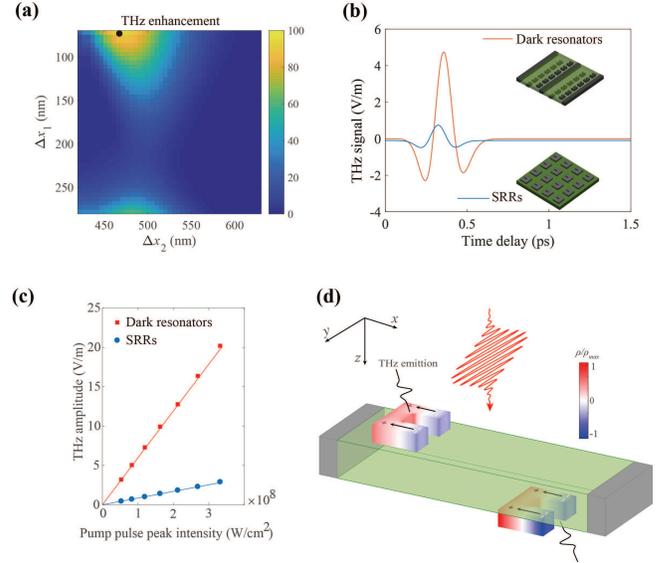}
\caption{\label{fig4} Terahertz enhancement of our design. (a) THz intensity enhancement map of our design. (b) Time trace of THz radiation from design (red) compared to that from traditional SRR metasurface (blue). (c) Terahertz signal strength in dependence of pump power for our design (red squares) and for traditional SRRs (blue dots). The straight lines are from linear fitting. (d) Distributions of DFG currents (black arrows) and charge density (color plot) of the SRRs. ``$+$" and ``$-$" indicate charge accumulations.}
\end{figure}

For a rigorous analysis of THz radiation from the metasurface, we performed a self-consistent FDTD calculation based on the hydrodynamic model and Maxwell’s equations (see Supplemental Material). A temporal optical pulse was applied, $E_y(t) = E_0{\rm{exp}}[-2{\rm{ln}}(2)(t-t_0)^2/\tau^2]{\rm{cos}}(\omega_0 t)$, where the driving frequency equals the resonant frequency of the dark mode, i.e., $\omega_0 = \omega_{\rm{dark}}$, the pulse duration $\tau = 140$ fs (spectral width $\sim$3.2 THz) and peak amplitude $E_0 = 2\times10^7$ V/m (see Supplemental Material for our estimation to thermal damage threshold of the metasurface). In our calculations, both linear and nonlinear dynamics of the electron gas under the infrared light excitation have been fully considered. In Fig.~\ref{fig4}(a), we show a performance map of our proposed THz emitter in term of THz enhancement defined as $\eta_{\rm{dark}}/\eta_{\rm{SRR}}$, where $\eta_{\rm{dark}}$ and $\eta_{\rm{SRR}}$ represent the THz conversion efficiency of our system and that of the traditional SRR metasurface \cite{Luo2014} (as a reference), respectively. By properly settling the two SRRs, we can significantly eliminate the radiation damping, resulting in more than 120-fold enhancement in THz emission power compared to the reference case. Note that the THz enhancements for different configurations are inversely proportional to the linear radiation damping as shown in Fig.~\ref{fig4}(a). Figure~~\ref{fig4}(b) gives typical time-domain trace detected at the receiving port for our metasurface THz emitter compared to that for the reference SRR case. Upon increasing the pump pulse intensity, the generated THz signal amplitude increases linearly as presented in Fig.~\ref{fig4}(c), and our design does show a dramatic emission enhancement over the analog dependence for the conventional SRR metasurface under the same conditions. In Fig.~~\ref{fig4}(d), we illustrate the charge density distribution (marked in color) and the currents (indicated with arrows) in the two SRRs at the maximum THz conversion efficiency. Both SRRs, possessing the same polarization with negligible phase difference, contribute to the THz emission (polarized along $x-$axis) coherently, the superposition of which leads to the total enhanced THz radiation, emitting perpendicular to the metasurface on both sides (see Supplemental Material). Moreover, the radiated THz signal polarized in the \textit{x}-direction can be written in terms of second-order nonlinear susceptibility $\chi^{(2)}$ of the metasurface as $E_x^{\rm{THz}}(\omega) \sim \chi^{(2)}\omega^2 e^{-\omega^2\tau^2} \Rightarrow E_x^{\rm{THz}}(t) \sim \chi^{(2)}(1-2t^2/\tau^2)e^{-t^2/\tau^2}/\tau^2$.


The THz signal depends on the temporal envelope of the incident pulse, and THz spectrum has the peak at $\omega=2/\tau$ with a bandwidth $\Delta\omega_{\rm{FWHM}}\approx2.31/\tau$. From the field of second harmonic radiation, we evaluated the effective nonlinear susceptibility of our designed metasurface $\chi^{(2)}_{\rm{dark}}\approx 2.4\times10^{-11}$ m/V, while that of the traditional SRR metasurface $\chi^{(2)}_{\rm{SRR}} \approx 1.7\times10^{-12}$ m/V (consistent with the reported $1.6\times10^{-12}$ m/V in \cite{Luo2014}).The discussions to the case with substrate can be found in Supplemental Material. In the view of potential experiments, conversion efficiency is an important parameter for an intuitive evaluation on nonlinear energy extraction. In practice, for a specific experimental setup, it depends on many factors, such as pump power, pump polarization, pulse duration, incidence and focusing conditions, phase-matching, etc. Our theoretical work described here resolves a fundamental problem obstructing the improvement of THz generation efficiency with nonlinear metasurfaces, and to understand the fundamental physics, we considered a simplified model of a spatially infinite surface under plane-wave illumination, so the model is less suitable to predict practical absolute numbers as would be observed in a particular experiment. Therefore, we conducted a side-by-side comparison between the conventional, not radiation-managed SRR metasurface \cite{Luo2014}, for which experimental data is available, and our configuration, which eliminates linear radiation damping, within the same numerical simulations. It is revealed our scheme has $\chi^{(2)}$ more than one order of magnitude larger than that of the regular SRR metasurface, and consequently would generate over two-order enhanced THz emission under the same conditions. On the other hand, Ref. [35] reported that the SRR metasurface experimentally showed about the same order of THz generation as a 0.2 mm-thick ZnTe crystal, so our proposed configuration should show a much better performance in generating broadband THz emission in real experiments.

In conclusion, our proposed system demonstrates effective suppression of radiation damping and extremely high-efficiency THz generations with a specifically designed metasurface. Such significant improvements in nonlinear conversion efficiency were made possible by our judiciously designed dark-element-integrated perfect absorbing sheet, which essentially tackled the common efficiency limit in two crucial aspects, i.e., eliminating radiative damping of linear scatterings and enhancing local field for nonlinear signals. Although our design presented in this Letter aims for efficient and compact THz generations, the new concept is general and applicable to other nonlinear processes. We believe the flexibilities in tailoring meta-atoms of metamaterials specially metasurfaces will open up new opportunities for THz technologies development and versatile nanoscale nonlinear devices.

\begin{acknowledgments}
Work at Ames Laboratory was supported by the US Department of Energy, Office of Basic Energy Science, Division of Materials Science and Engineering (Ames Laboratory is operated for the US Department of Energy by Iowa State University under contract No. DE-AC02-07CH11358). Work at FORTH was supported by the European Research Council under the ERC Advanced Grant No. 320081 (PHOTOMETA). The work was also supported by NSFC (Nos. 61601166, 61701001,61701003), National Natural Science Fund for Excellent Young Scholars (No. 61722101), Universities Natural Science Foundation of Anhui Province (Nos. KJ2017ZD51 and KJ2017ZD02), and Thousand Talents Program for Distinguished Young Scholars of China. The authors would like to thank Peng Zhang, Liang Luo and Xu Yang for fruitful discussions.
\end{acknowledgments}
%
%

\bibliographystyle{apsrev4-1}
\bibliography{ref_back}

\end{document}